\documentclass[aip,graphicx,submission,preprint]{revtex4-1}
\usepackage{graphicx}

\begin{document}

\title{Hot electron bolometer heterodyne receiver with a 4.7-THz quantum cascade laser as a local oscillator} 



\author{J. L. Kloosterman}
\email[]{jlkloost@email.arizona.edu}
\affiliation{Department of Electrical and Computer Engineering, 1230 E. Speedway Blvd., University of Arizona, Tucson, AZ 85721 USA}

\author{D. J. Hayton}
\affiliation{SRON Netherlands Institute for Space Research, Groningen/Utrecht, Netherlands}

\author{Y. Ren}
\affiliation{Kavli Institute of NanoScience, Delft University of Technology, Lorentzweg 1, 2628 CJ Delft, The Netherlands}
\affiliation{Purple Mountain Observatory (PMO), Chinese Academy of Sciences, 2 West Beijing Road, Nanjing, JiangSu 210008, China, and Graduate School, Chinese Academy of Sciences, 19A Yu Quan Road, Beijing 100049, China}

\author{T. Y. Kao}
\affiliation{Department of Electrical Engineering and Computer Science, Massachusetts Institute of Technology (MIT), Cambridge, Massachusetts 02139, USA}

\author{J. N. Hovenier}
\affiliation{Kavli Institute of NanoScience, Delft University of Technology, Lorentzweg 1, 2628 CJ Delft, The Netherlands}

\author{J. R. Gao}
\thanks{j.r.gao@tudelft.nl}
\affiliation{SRON Netherlands Institute for Space Research, Groningen/Utrecht, Netherlands}
\affiliation{Kavli Institute of NanoScience, Delft University of Technology, Lorentzweg 1, 2628 CJ Delft, The Netherlands}

\author{T. M. Klapwijk}
\affiliation{Kavli Institute of NanoScience, Delft University of Technology, Lorentzweg 1, 2628 CJ Delft, The Netherlands}

\author{Q. Hu}
\affiliation{Department of Electrical Engineering and Computer Science, Massachusetts Institute of Technology (MIT), Cambridge, Massachusetts 02139, USA}

\author{C. K. Walker}
\affiliation{Steward Observatory, 933 N Cherry Ave., Rm N204, University of Arizona, Tucson, AZ 85721 USA}

\author{J. L. Reno}
\affiliation{Sandia National Laboratories, Albuquerque, NM 87185-0601, USA}


\date{\today}

\begin{abstract}
We report on a heterodyne receiver designed to observe the astrophysically important neutral atomic oxygen [OI] line at 4.7448 THz.  The local oscillator is a third-order distributed feedback Quantum Cascade Laser operating in continuous wave mode at 4.741 THz. A quasi-optical, superconducting NbN hot electron bolometer is used as the mixer.  We recorded a double sideband receiver noise temperature (T$^{DSB}_{rec}$) of 815 K, which is $\sim$7 times the quantum noise limit ($\frac{\rm{h}\nu}{2\rm{k_B}}$) and an Allan variance time of 15 s at an effective noise fluctuation bandwidth of 18 MHz.  Heterodyne performance was confirmed by measuring a methanol line spectrum.
\end{abstract}


\maketitle 


Astronomers have long been interested in the fine structure line of [OI] at 4.7448 THz. [OI] probes the star formation process and is the most important cooling line in the interstellar medium (ISM)\citep{tielens85} for gas clouds with densities of n $> 10^4$ cm$^{-3}$.  Large scale surveys with extremely high spectral resolution and sensitivity are required to disentangle large scale kinematics and energetics within these clouds. Such high spectral resolution observations require the development of super-THz ($>$ 3 THz) heterodyne receivers.  Due to strong water absorption in the atmosphere, it is not possible to observe the [OI] line from ground-based telescopes.  Therefore, an astronomical [OI] receiver requires a compact local oscillator (LO) that can be readily integrated into space-based or suborbital observatories.

There are several candidate THz technologies for use in LO systems. These technologies include Schottky diode based multiplier chains, gas lasers, and quantum cascade lasers (QCLs).  QCL’s are currently the only technological approach that leads to devices small and powerful enough \citep{kohler02} to be used in a variety of space-based, super-THz applications.  Furthermore, THz QCLs operating in CW mode have yielded line widths of $\sim$100 Hz \citep{vitiello12}, excellent power stability \citep{gao05}, and output powers over 100 mW \citep{williams05}, making them well-suited for high resolution spectroscopy.  Much progress has been made toward overcoming the challenges associated with using a QCL as an LO.  Frequency stabilization without the need of another THz source has been achieved using an absorption line within a methanol gas cell \citep{richter10b, ren12}.  Diverging far-field beam patterns and mode selectivity have been improved by using a 3$^{\rm{rd}}$-order distributed feedback (DFB) grating \citep{amanti10,kao12}.  

In this letter we report on a full demonstration of a heterodyne receiver using a THz quantum cascade laser as local oscillator.  In contrast to previous publications \citep{gao05,ren11}, significant progress has been made on DFB QCLs by changing the tapered corrugations.  At 4.7 THz, this QCL is the highest frequency ever reported using the 3$^{\rm{rd}}$-order DFB structure.  Furthermore, by introducing an array of 21 DFB lasers with a linear frequency coverage and a 7.5 GHz frequency spacing, we can target a specific LO frequency.  An unprecedented high sensitivity for a heterodyne receiver was measured at 4.741 THz along with a 15 s Allan variance time, the first time such stability has been reported with this combination. Lastly, a theoretical model for methanol molecular lines has been verified at 4.7 THz, which was not possible until now.

\begin{figure}
\includegraphics[scale=1.1]{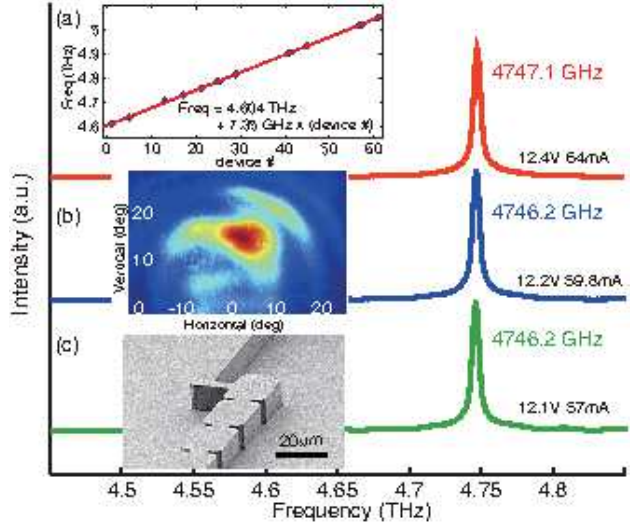}
\caption{(Color online) CW spectra of a 4.7 THz QCL (at 77 K)  measured at different bias voltages.  (a) Frequencies of an array with 11 devices in pulse mode (at 10 K) demonstrating the frequency selectivity of a 21-element third-order DFB array. (b) Beam pattern of the QCL. (c) Scanning Electron Microscope (SEM) image for a taper-horn third-order DFB laser. The contact pad connects to the side of the last period of the DFB grating. }\label{qcl}
\end{figure}

The THz QCL active region is based on a four-well resonant-phonon depopulation design in a metal–-metal waveguide.  Cavity structure with a lateral corrugated third-order DFB grating similar to those demonstrated in Amanti et al. \citep{amanti10} were used to provide frequency-selectivity and also to the improve far-field beam pattern.  We improve upon this design by changing the shape of the corrugated gratings from a traditional square tooth to a tapered shape as shown in Fig. \ref{qcl}c.  According to an electromagnetic finite-element (FEM) simulation, the taper-horn shape increases the radiation loss from the unwanted upper band-edge mode while marginally reducing the radiation loss for the desired third-order DFB mode, hence improving mode selectivity in order to ensure a robust single-mode operation.  Effectively, this approach leverages a trade-off between the output power efficiency and mode discrimination.  With this improved frequency selectivity, we realize a linear frequency coverage of 440 GHz, from 4.61 to 5.05 THz as shown in Figure \ref{qcl}a with robust single-mode operation on the same gain medium.  These grating periods range from 28.5 to 25 $\mu$m, which cover $\sim$80\% of the gain spectrum.

The third-order DFB QCL arrays were fabricated using standard metal–-metal waveguide fabrication techniques, contact lithography, and inductively-coupled–-reactive ion etching (ICP-RIE) to define the laser mesas with the Ti/Au top contact acting as the self-aligned etch mask.  A 300 nm SiO$_2$ electrical insulation layer was used for the isolation of the contact pads. Each array consists of 21 DFB lasers arranged in a similar manner as in Lee et al. \citep{lee12} with a $\sim$7.5 GHz frequency spacing.  The position where the contact pads connect to the DFB laser was chosen to minimize unwanted perturbation to the grating boundary condition.

The device used in the heterodyne experiment has a width of 17 $\mu$m and 27 grating periods with an overall device length of $\sim$0.76 mm.  The measured CW output power is 0.25 mW with $\sim$0.7 W DC of power dissipation at 10 K and a main beam divergence of $\sim$12$^{\circ}$, as shown in Figure \ref{qcl}b. CW lasing at 4.7471 THz is realized at 77 K with a 12.4 V bias voltage, which is within 3 GHz of the target [OI] line (see Fig. \ref{qcl}a).  For the heterodyne measurement described below, the device is operated at $\sim$10 K.

HEBs are the preferred mixer for frequencies above 1.5 THz and have been used up to 1.9 THz in the {\it Herschel Space Telescope} \citep{chered08} and the {\it Stratospheric THz Observatory} \citep{walker10}, and up to 2.5 THz in the {\it Stratospheric Observatory For Infrared Astronomy} \citep{putz12,heyminck12}.  We use a NbN HEB mixer, which was developed by SRON and TU Delft.  Nb contact pads connect a 2 $\times$ 0.2 $\mu$m$^2$ superconducting bridge to a tight winding spiral antenna, which is suitable for super-THz frequencies.  With the application of both electrical bias and optical pumping from an LO source, a temperature distribution of hot electrons is maintained producing a resistive state in the center of the bridge.  Incoming signals modulate temperature causing a modulation in the resistance to create heterodyne mixing \citep{barends05}.

\begin{figure}
\includegraphics[scale=1]{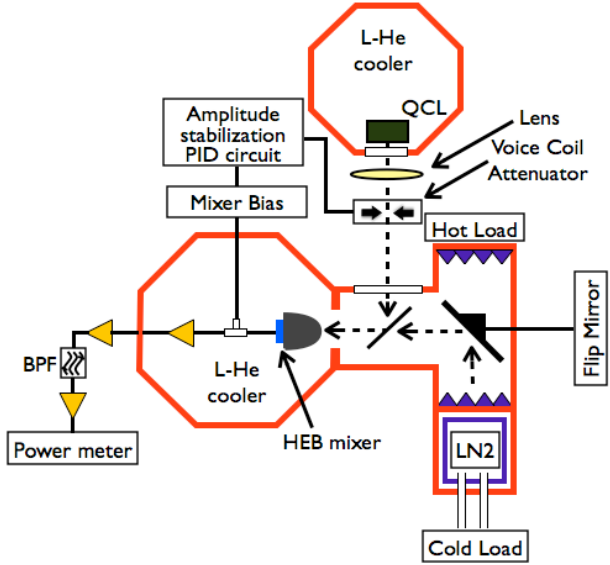}
\caption{(Color online) Laboratory setup for heterodyne QCL-HEB measurements.}\label{bd}
\end{figure}

The test setup for measuring T$^{DSB}_{rec}$ is shown in Fig. \ref{bd}.  The QCL was mounted in a liquid helium cryostat and operated at $\sim$10 K.  The beam was focused by an ultra-high molecular weight polyethylene (UHMW-PE) lens, which is $\sim$80\% transmissive at 4.7-THz.  A voice coil attenuator together with a proportional - integral - derivative (PID) feedback loop is used to stabilize the power output of the QCL during T$^{DSB}_{rec}$ and Allan variance measurements \citep{hayton12} (where noted).  In this case the HEB DC current is used as a power reference signal.  The beam entered a blackbody hot/cold vacuum setup attached to the HEB cryostat via a second UHMW-PE window and then was reflected by a 3 $\mu$m mylar beam splitter.  This cryostat was cooled to 4.2 K.  The HEB was mounted to the back side of a 10 mm Si lens with an anti-reflection coating designed for 4.25 THz.  The first stage low noise amplifier (LNA) was attached to the cold plate and operated at 4.2 K.  The LNA noise temperature was 3 K with a gain of 42 dB measured at 15 K.  Outside the dewar, room temperature amplifiers and an 80 MHz wide band pass filter (BPF) centered at 1.5 GHz were used to further condition the intermediate frequency (IF) signal before the total power was read using an Agilent E4418B power meter.

The total optical losses in the setup are $\sim$20 dB or about 99\% of the QCL emission, including the mylar beam splitter efficiency and atmospheric absorption in the optical path from the LO to the HEB mixer \citep{optlosses}.  Based on the QCL output power of 250 $\mu$W, a maximum of 2.5 $\mu$W can couple into the detector.  By using the isothermal method based on IV curves\citep{ekstrom95}, the maximum LO power recorded by the detector is $\sim$290 nW.  Thus, with a lens, 10\% to 15\% of the available power was coupled into the HEB.  Because of the beam pattern of the third-order DFB structure, this is considerably improved over the 1.4\% coupling efficiency of previous generation QCLs\citep{gao05}.

Receiver sensitivity was measured using the Y-factor method.  Eq. \ref{trecn} is used to convert a Y-factor to a T$^{DSB}_{rec}$. The Callen-Welton temperatures at 4.7 THz are T$_{\rm{eff,hot}} = 309$ K and T$_{\rm{eff,cold}} = 126$ K \citep{callen51}.

\begin{equation}\label{trecn}
\rm{T_{N,rec} = \frac{T_{eff,hot}-YT_{eff,cold}} {Y-1}}
\end{equation}

\begin{figure}
\includegraphics[scale=0.6]{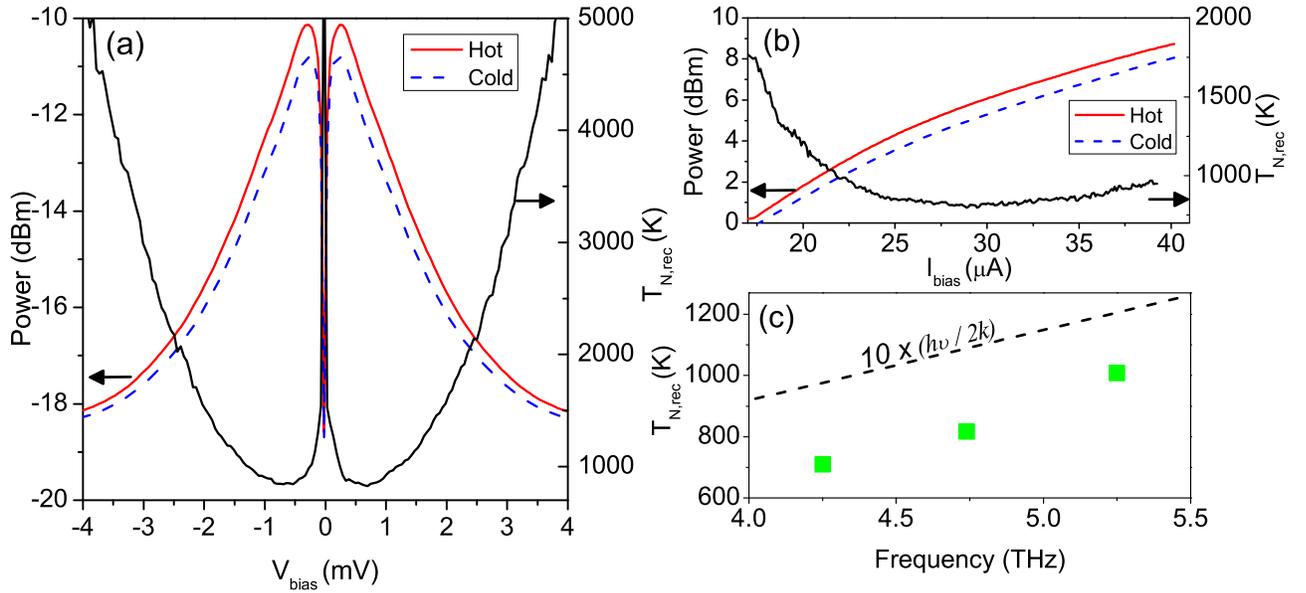}
\caption{(Color Online) (a) IF power measurements as functions of bias voltage with the calculated T$^{DSB}_{rec}$ plotted on the right hand side. (b) IF power measurements as functions of stabilized bias current with the calculated T$^{DSB}_{rec}$ plotted on the right hand side.  (c) T$^{DSB}_{rec}$ for 4.25, 4.74, and 5.25 THz using a 3 $\mu$m beam splitter.  A gas laser was used as an LO at 4.25 ad 5.25 THz and a QCL was used as an LO at 4.74 THz.  Ten times the quantum noise limit is also shown with the dashed line.}\label{trec}
\end{figure}

The Y-factor was measured using three different methods.  In Fig. \ref{trec}a the measured IF power was swept as a function of bias voltage with a fixed LO power. We corrected for direct detection by adjusting the LO power so that the IV curves were on top of one another.  The best T$^{DSB}_{rec}$ was found to be 825 K at a bias of 0.7 mV and 30 $\mu$A.  Recently it has become possible to accurately sweep LO power by attenuating the LO signal with a stabilized voice coil attenuator \citep{hayton12} and plot the resulting HEB bias current as a function of output power (see Fig. \ref{trec}b).  This method reduces direct detection and results in an average T$^{DSB}_{rec}$ of 810 K around a bias of 0.65 mV and 29 $\mu$A.  This current corresponds to 220 nW of LO power.  The third method, not shown, chops between hot and cold loads with a stabilized current.  It also produced a T$^{DSB}_{rec}$ of 810 K at the same operating point.  Thus we obtain a T$^{DSB}_{rec}$ of 815 K by averaging the three methods. This T$^{DSB}_{rec}$ is $\sim$7 times the quantum noise limit $\left( \frac {h\nu}{2 k_B} \right)$.  

To demonstrate the QCL adds no additional noise to the receiver system, T$^{DSB}_{rec}$ measurements were taken with a gas laser at 4.25 and 5.25 THz.  We recorded 750 K at 4.25 THz and 950 K at 5.25 THz with the same HEB receiver.  Fig. \ref{trec}c shows that all three T$^{DSB}_{rec}$ scale linearly with frequency.  This suggests that the QCL is a clean LO source.  These measurements improve upon the previously published T$^{DSB}_{rec}$ of 860 K at 4.25 THz and 1150 K at 5.25 THz\citep{zhang10}.  We attribute most of this ($\gtrsim$12\%) improvement to a new IF mixer circuit.

HEB receivers have been plagued by stability issues, which can now largely be attributed to instability in the received LO power at the detector \citep{hayton12}.  Allan variance measurements are important in determining the optimum integration time on a source between instrument calibrations.  For this purpose the noise temperature setup of Fig. \ref{bd} was modified to include a two-way power splitter at the end of the IF chain. Each of the output channels was then sent through a band pass filter, one centered at 1.25 GHz and the other at 1.75 GHz.  This enabled measurements of the Allan variance in the spectroscopic configuration (the spectral difference between the two channels), which yields greater Allan variance times because it effectively filters out longer period baseline variations.

\begin{figure}
\includegraphics[scale=0.4]{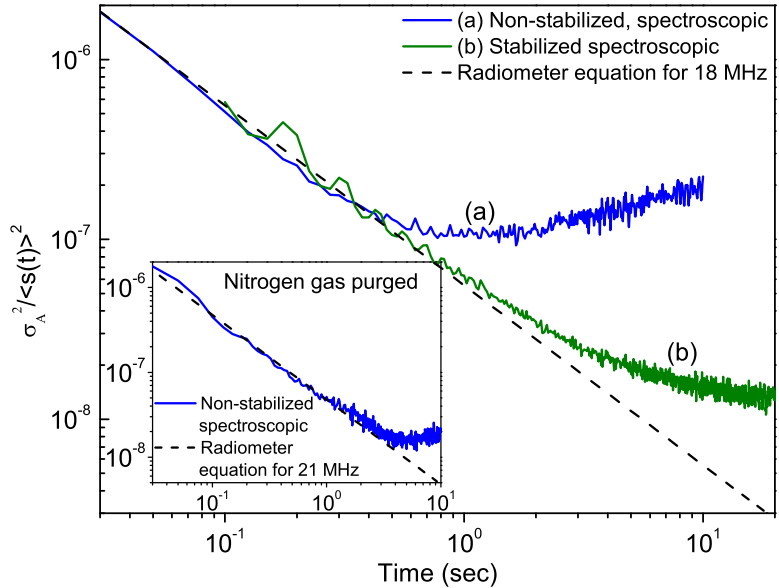}
\caption{(Color online) Measurements in air for (a) non-stabilized spectroscopic and (b) stabilized spectroscopic Allan variance curves.  In the inset is an Allan variance curve for a non-stabilized spectroscopic measurement taken with the air purged by nitrogen gas.}\label{allan}
\end{figure}

The results from our receiver are shown in Fig. \ref{allan}.  We found that the non-stabilized Allan variance time was $\sim$1 s and the stabilized Allan variance time was $\sim$ 15 s with an 18 MHz noise fluctuation bandwidth.  The resulting Allan variance time from a shorted IF chain was sufficiently long enough to eliminate the IF chain as a source of instability.  Next, we purged the air between the QCL window and vacuum setup with nitrogen gas.  This improves the non-stabilized, spectroscopic Allan time to $\sim$7 s as shown in the inset of Fig \ref{allan}, suggesting that atmospheric turbulence at 4.7 THz may be a large contributor to the instability in the system.

\begin{figure}
\includegraphics[scale=0.4]{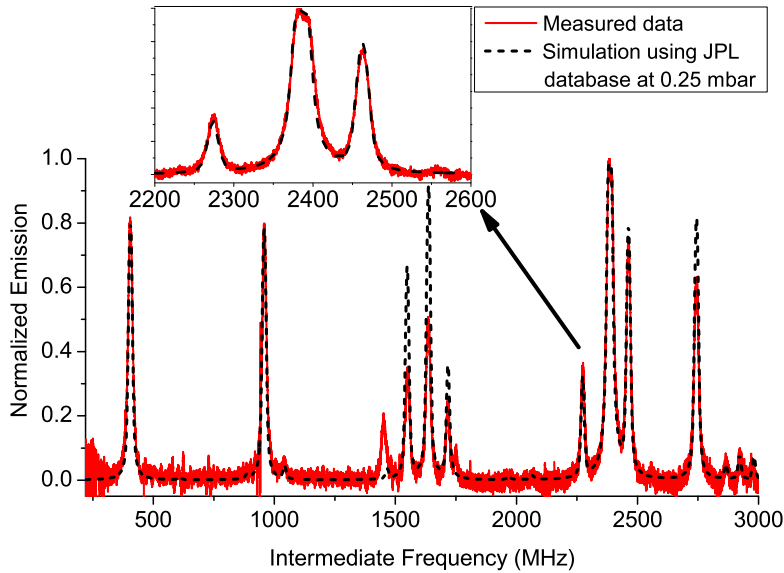}
\caption{(Color online) A DSB spectrum of methanol with an LO frequency of 4.74093 THz compared with the predicted spectrum from the JPL catalog.}\label{spectrum}
\end{figure}

In order to demonstrate the functionality of the receiver for heterodyne spectroscopy, the receiver was used to measure a spectrum of methanol gas (CH$_3$OH).  A methanol gas cell was attached to an external input port on the hot/cold vacuum setup so that there was no air in the signal path.  The QCL was operated at a bias voltage of 11.8 V.  The results, averaged over 18 s of integration time, are shown in Fig. \ref{spectrum} along with a simulation \citep{ren10c} at 0.25 mbar that predicts line widths based on the frequencies and line strengths from the JPL spectral catalog \citep{pickett98,xu08}.  The lines from 1500-1700 MHz are attenuated because the FFTS upper band (1500-3000 MHz) high pass filter has a cut-off frequency of 1700 MHz.  The best-fit frequency for the QCL is 4.740493 THz, which is close to the HEB bandwidth ($\sim$4 GHz) for the [OI] line.  The verification of the JPL spectral catalog is also important for the frequency locking of the QCL \cite{ren12}.

In conclusion, we have demonstrated a 4.7-THz HEB-QCL receiver with a measured sensitivity of 815 K and spectroscopic Allan time of 15 s. This T$^{DSB}_{rec}$ is 85 times lower than a previous Schottky receiver \citep{boreiko96}. Heterodyne performance was verified by observing a methanol spectrum. The performance of this receiver indicates THz receiver technology has reached a level of maturity that will permit large-scale [OI] surveys of the interstellar medium to take place, such as those planned by the Gal/Xgal Ultra-Long Duration Spectroscopic Stratospheric Terahertz Observatory (GUSSTO). 

We acknowledge G. Goltsman’s group at MSPU for the provision of NbN films.  We would like to thank John C. Pearson for his help in understanding methanol lines in the JPL catalog near 4.7 THz.  The work of the University of Arizona was supported by NASA grant NN612PK37C.  The work in the Netherlands is supported by NWO, KNAW, and NATO SFP.  The work at MIT is supported by NASA and NSF. The work at Sandia was performed, in part, at the Center for Integrated Nanotechnologies, a U.S. Department of Energy, Office of Basic Energy Sciences user facility. Sandia National Laboratories is a multiprogram laboratory managed and operated by Sandia Corporation, a wholly owned subsidiary of Lockheed Martin Corporation, for the U.S. Department of Energy‚ National Nuclear Security Administration under contract
DE-AC04-94AL85000.


\begin{thebibliography}{25}
\bibitem [{opt()}]{optlosses}%
  \BibitemOpen
  \href@noop {} {}\bibinfo {note} {The optical losses are due primarily to the
  UHMW-PE lens and cryostat windows ($\sim$3 dB), air ($\sim$3.5 dB), mylar
  beam splitter ($\sim$9 dB), QMC IR filter ($\sim$0.8 dB), coated Si lens
  ($\sim$1 dB), and spiral antenna ($\sim$3 dB).}\BibitemShut {Stop}%
\end{thebibliography}
%

\end{document}